\documentclass[aps,pra,twocolumn,showpacs,superscriptaddress]{revtex4-1}  

\usepackage{graphicx}  
\usepackage{dcolumn}   
\usepackage{bm}        
\usepackage{amssymb}   
\usepackage{amsmath}

\usepackage[colorlinks=true,breaklinks=true,allcolors=blue]{hyperref}
\usepackage{url}
\usepackage{txfonts}

\usepackage{graphics}
\usepackage{pstricks}
\usepackage{tikz}
\usepackage{graphicx}
\usepackage{color}
\usepackage{hyperref}

\renewcommand{\vec}[1]{\mathbf{#1}}
\newcommand{\eq}[1]{(\ref{eq:#1})}
\newcommand{\Eq}[1]{Eq.\,\eqref{eq:#1}}

\newcommand{\Fig}[1]{Fig.~\ref{fig:#1}}

\newcommand{\Sect}[1]{Sect.~\ref{sec:#1}}

\newcommand{\App}[1]{App.~\ref{app:#1}}

\makeatletter
\let\cat@comma@active\@empty
\makeatother

\begin{document}


\title{Violation of single-length scaling dynamics via spin vortices in an isolated spin-1 Bose gas}
\author{C.-M. Schmied}
\affiliation{Kirchhoff-Institut f\"ur Physik,
             Ruprecht-Karls-Universit\"at Heidelberg,
             Im~Neuenheimer~Feld~227,
             69120~Heidelberg, Germany}
\affiliation{Department of Physics, 
		Dodd-Walls Centre for Photonic and Quantum Technologies, 
		University of Otago, 
		Dunedin 9016, New Zealand}
\author{T. Gasenzer}
\affiliation{Kirchhoff-Institut f\"ur Physik,
             Ruprecht-Karls-Universit\"at Heidelberg,
             Im~Neuenheimer~Feld~227,
             69120~Heidelberg, Germany}
\author{P. B. Blakie}
\affiliation{Department of Physics, 
		Dodd-Walls Centre for Photonic and Quantum Technologies, 
		University of Otago, 
		Dunedin 9016, New Zealand}
\date{\today}

\begin{abstract}
We consider the phase ordering dynamics of an isolated quasi-two-dimensional spin-1 Bose gas quenched into an easy-plane ferromagnetic phase.  
Preparing the initial system in an unmagnetized anti-ferromagnetic state the subsequent ordering involves both polar core and Mermin-Ho spin vortices, with the ratio between the different vortices controllable by the quench parameter.
Ferromagnetic domain growth occurs as these vortices annihilate. 
The distinct dynamics of the two types of vortices means that the domain growth law is determined by two macroscopic length scales, violating the standard dynamic scaling hypothesis.  
Nevertheless we find that universality of the ordering process manifests in the decay laws for the spin vortices.
\end{abstract}

\pacs{%
11.10.Wx 		
03.75.Lm 	  	
47.27.E-, 		
67.85.De 		
}

\maketitle

\section{Introduction}
\label{sec:Introduction}
Quenching a many-body system from a disordered into an ordered phase leads to the formation of spatial field patterns of linear and non-linear field excitations, including solitonic waves, (quasi)topological defects, domain walls and more irregular structures \cite{cross1993a, Hohenberg1977a,Bray1994a.AdvPhys.43.357, Henkel2010a.NonEqPhaseTransitions2}. 
These patterns will consecutively start to grow developing macroscopic order in the system. 
The pattern size is generically given by a characteristic length scale $L$, which is initially set by the quench and grows in the course of the ordering process. 
Once it exceeds characteristic microscopic length scales, the phase ordering typically becomes universal such that it exhibits a power-law growth in time as $L(t) \sim t^{\, \beta}$, with universal scaling exponent $\beta$.

Due to their rich phase diagram \cite{Stenger1999a, Ho1998a} and their high controllability in experiments, spinor Bose gases are ideally suited for studying universal dynamics in quantum many-body systems.
Apart from experimental studies of short-time dynamics following quenches between different phases \cite{Sadler2006a,  Bookjans2011a}, subsequent domain coarsening of spin textures, without reference to universal scaling, has been observed in the long-time dynamics of a quasi-2D spin-1 system \cite{Guzman2011a}.
Universal scaling with exponent $\beta \simeq 1/2$ has recently been observed experimentally in a ferromagnetic spin-1 Bose gas in a near-1D geometry \cite{Prufer:2018hto}. 
Theoretical studies have shown that universal scaling can occur in the ordering process of one- and  quasi two-dimensional (quasi-2D) spin-1 as well as binary Bose gases after a parameter quench into an ordered phase \cite{Fujimoto2018a, Fujimoto2018b, SchmiedPhysRevA.99.033611, Williamson2016a.PhysRevLett.116.025301, Williamson2016a.PhysRevA.94.023608, Williamson2017a.PhysRevLett.119.255301, Symes2017a,Bourges2016a.arXiv161108922B.PhysRevA.95.023616, Hofmann2014PhRvL.113i5702H}.  

Phase ordering kinetics and coarsening are commonly discussed in dissipative systems \cite{Bray1994a.AdvPhys.43.357,Bray2000PhRvL..84.1503B,Rutenberg1995b,Rutenberg1995a.PhysRevLett.74.3836,Blundell1994,
PhysRevB.41.6724,Rojas1995a,Kudo2013a}, where the universal ordering process is characterized by the underlying dynamics of (quasi)topological excitations.
Dissipative coarsening forms a special case of more general spatio-temporal universal dynamical phenomena far from equilibrium which 
can occur in both, open and isolated (quantum) many-body systems \cite{Taeuber2014a.CriticalDynamics,Tauber2013a,Henkel2010a.NonEqPhaseTransitions2,Schmied:2018mte}.
Following a quench far out of equilibrium, a system can in general approach a non-thermal fixed point \cite{Berges:2008wm,Berges:2008sr,Scheppach:2009wu,Orioli:2015dxa, Chantesana:2018qsb.PhysRevA.99.043620,Prufer:2018hto,Erne:2018gmz}.
Such fixed points have been discussed and experimentally observed with \cite{Nowak:2010tm,Nowak:2011sk,Schole:2012kt,Karl:2013kua,Karl2017b.NJP19.093014,Erne:2018gmz}  and without \cite{Berges:2008wm,Berges:2008sr,Scheppach:2009wu,Berges:2010ez,Berges:2013fga,Orioli:2015dxa,Berges:2015kfa,Prufer:2018hto} reference to ordering patterns and kinetics as well as topological defects.

Here, we numerically study phase ordering dynamics in an isolated quasi-2D spin-1 Bose gas after a quench into the easy-plane ferromagnetic phase. Our key observation is that by initializing the system in an unmagnetized anti-ferromagnetic state (see, e.g., \cite{Seo2015a}) the initial quench dynamics 
produces both polar-core vortices (PCVs) and Mermin-Ho vortices (MHVs) as the ordered domains develop. 
These two types of spin vortices have distinct universal decay laws, with the respective vortex densities introducing two independent macroscopic length scales.
The ratio of these vortices can be engineered by the quadratic Zeeman energy, 
which we find to have a striking effect on the time scales and nature of the ordering. 
Earlier work studying spin-1 quench dynamics found that only PCVs played a role in the phase ordering \cite{Williamson2016a.PhysRevLett.116.025301}. 
Phase ordering of a (non-quenched) spin-1 system containing only MHVs was investigated in Ref.~\cite{Kudo2015a}, with the initial vortices inserted randomly into the initial equilibrium state.

Our paper is organized as follows: In \Sect{Spin1BoseGas}, we introduce the model under consideration as well as the applied quench protocol. 
In \Sect{SpinVortices}, we review the relevant spin vortices of the system in the easy-plane ferromagnetic phase.
The main results are presented in \Sect{PhaseOrderingDynamics}. We begin with a brief discussion of the numerical methods used to simulate the dynamics after the quench. We then present an algorithm for the detection of different types of spin vortices. With this at hand we examine the phase ordering dynamics of the system. We extract the universal decay laws of the spin vortices and show the violation of single-length scaling.
We finally draw our conclusions and give an outlook to future work in \Sect{Conclusion}.

\section{Spin-1 Bose gas} 
\label{sec:Spin1BoseGas}
We consider a homogeneous quasi-2D spin-1 Bose gas described by the Hamiltonian \cite{Stamper-Kurn2013a.RevModPhys.85.1191}
\begin{equation}
H = \int d^2 x \left [ \bm{\psi}^{\dagger} \left( -\frac{\hbar^{2}\nabla^2}{2M}  + q f_z^2 \right) \bm{\psi}+  \frac{c_0}2  n^2 + \frac{c_1}2  \lvert  \vec{F} \rvert ^2 \right],
\label{eq:Hamiltonian}
\end{equation}
where $\bm{\psi} = \left(\psi_1, \psi_0, \psi_{-1} \right)^T$ is the bosonic spinor field whose components account for the magnetic sublevels $m_F = 0, \pm 1$ of the $F=1$ hyperfine manifold. The quadratic Zeeman energy
$q$  along the $z$-direction can be controlled by external magnetic fields. 
We work in a frame where a possible homogeneous linear Zeeman shift has been absorbed into the definition of the fields. 
Spin-independent contact interactions are described by the term $c_0 n^2$, where $n = \bm{\psi}^{\dagger} \bm{\psi}\equiv \sum_{m}\vec{\psi}_{m}^{\dagger} \vec{\psi}_{m}$ is the total density. 
Spin-dependent interactions are characterized by the term $c_1 \lvert \vec{F} \rvert^2$, where $\vec{F} = \bm{\psi}^{\dagger} \vec{f} \bm{\psi}$ is the spin density and $\vec{f} = \left (f_x, f_y,f_z \right)$ is a vector that contains the matrices forming the fundamental representation of the spin-1 algebra. 
This term includes the redistribution of population between the three components via spin-mixing dynamics \cite{Stamper-Kurn2013a.RevModPhys.85.1191}.

For   ferromagnetic interactions  (i.e.~$c_1 < 0$) and $q>0$ the system exhibits two different phases separated by a quantum phase transition (QPT)  \cite{Kawaguchi2012a.PhyRep.520.253}. For 
$q>  q_0 = 2 n_0 \lvert c_1 \rvert$ the system is in the polar phase where the mean-field ground state is unmagnetized and given by the state vector 
$\bm{\psi}_{\mathrm{P}} = e^{i \theta} \sqrt{n_0} \left (0,1,0 \right)^T$.
Here, $n_0$ is the homogeneous condensate density and  $\theta$ is a global phase distinguishing different realizations of the spontaneous symmetry breaking. 
For $0 < q <  q_0$  
the system is in the easy-plane ferromagnetic phase in which the mean-field ground state reads
\begin{equation}
  \bm{\psi}_{\mathrm{EP}} =  \sqrt{n_0} 
  \frac{e^{i \theta}} 2  
  \begin{pmatrix}
  e^{-i \phi}\sqrt{ 1 - q/q_0}   \\ \sqrt{  2 (1 + q/q_0) }  \\ e^{i \phi} \sqrt{ 1 - q/q_0}  
  \end{pmatrix},
\end{equation}
where $\phi$ denotes the angle of the in-plane magnetization with respect to the $x$-axis. 
The magnitude of the magnetization is 
$ \lvert F_{\perp}\rvert =  n_0   [1- (q/q_0)^{2}  ] ^{\frac{1}{2}}$.
We take $F_{\perp} = F_x + i F_y $ as the spin order parameter for this phase.

Beginning from a polar condensate  at $q>q_0$ it has been shown that after a sudden quench across the QPT  the system undergoes phase ordering dynamics within the easy-plane ferromagnetic phase  \cite{Williamson2016a.PhysRevLett.116.025301, Prufer:2018hto, SchmiedPhysRevA.99.033611}. 
In this work we investigate the phase ordering dynamics occurring  when quenching the system into the easy-plane ferromagnetic phase starting from the state
\begin{equation}
\bm{\psi}_{\mathrm{AF}} =  \sqrt{\frac {n_0} 2}
\begin{pmatrix}
e^{i \phi_1} \\ 0 \\  e^{i \phi_{-1}},
\end{pmatrix},
\label{EqAF}
\end{equation}
where $\phi_{\pm 1}$ denote  arbitrary phases.
This unmagnetized state is a mean-field ground state for the case of anti-ferromagnetic interactions ($c_1>0$), commonly referred to as the anti-ferromagnetic phase.
It can be easily generated experimentally by applying a $\pi/2$ rf-rotation to a polar condensate \cite{Seo2015a}.

Quenching into the easy-plane ferromagnetic phase gives the system an excess energy (relative to the easy-plane ground state) of
$\Delta \epsilon_{\mathrm{AF}} = q +  \tfrac{1}{4}q_0 \left( 1- q/q_0 \right)^2$.
This is larger than the excess energy for a polar initial condition ($\Delta \epsilon_{\mathrm{P}} =  \Delta \epsilon_{\mathrm{AF}} - q$), indicating that more heating will occur for the initial condition we employ. However, the degree of extra heating will be less for smaller $q$ and here we focus on the regime $q \lesssim 0.3 \, q_0$.

\begin{figure*}
\includegraphics[width = 0.92\textwidth]{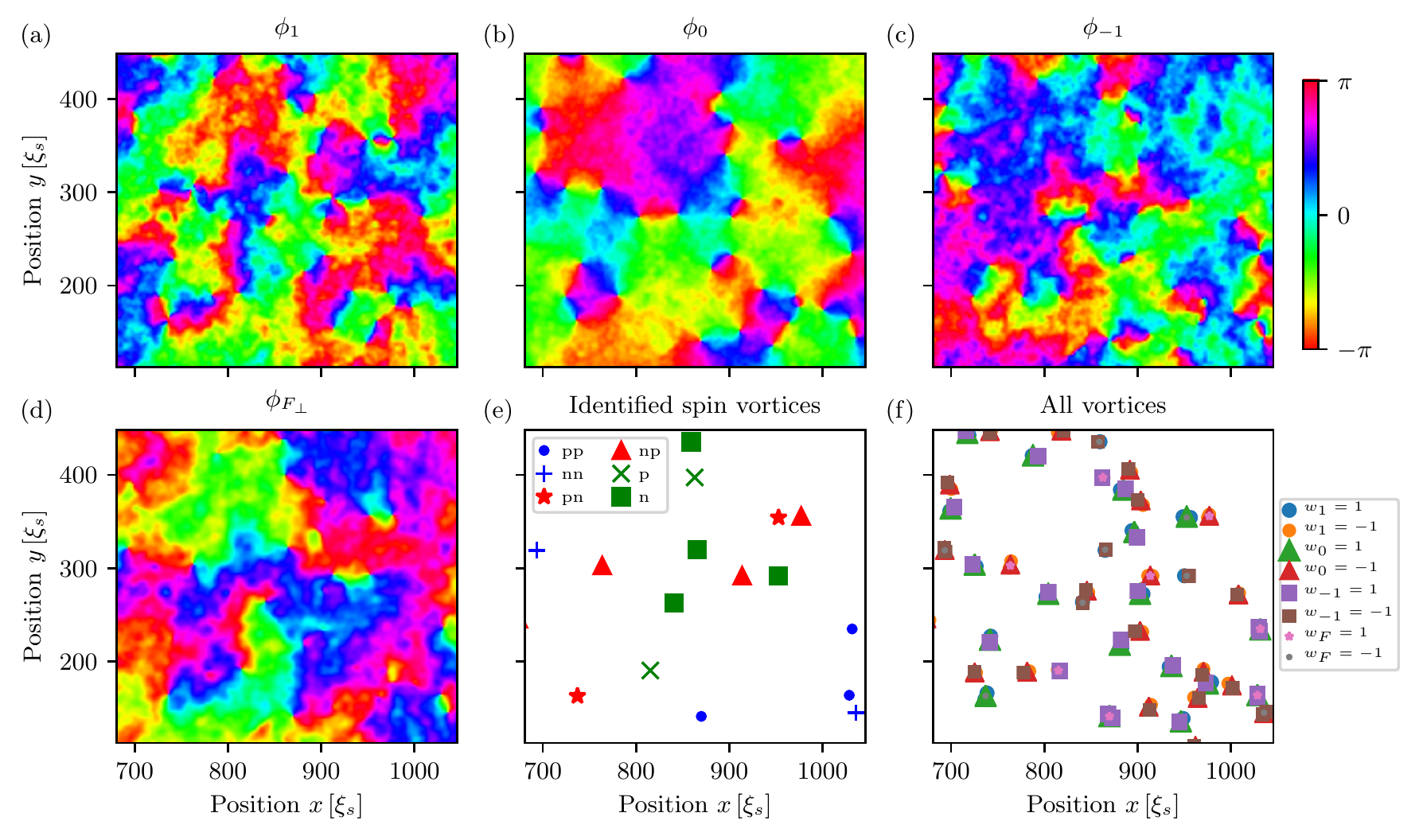}
\caption{\label{fig:SnapshotSingleRun} Image of a $336 \, \xi_s\times336 \, \xi_s$ subregion of the system at time $t = 2485 \, t_s$, where  $t_s = \hbar/q_0$ is the characteristic spin time,  
 for a quench to $ q = 0.15 \, q_0$. (a)-(c) Phase profiles $\phi_{m_F}=\mathrm{Arg}(\psi_{m_F})$ of the three $m_{\mathrm{F}}$ components. (d) Phase field of the transversal spin revealing the spin circulations. (e) Identified spin vortices: Vortex-antivortex pairs are shown with the same color code. All  types of Mermin-Ho vortices (MHVs) (blue pluses/dots and red stars/triangles) and polar core vortices (PCVs) (green crosses/squares) are present. Vortex labeling and details of the identification procedure are given in the main text.
(f) All vortices detected in the data depicted in (a)-(d) including their winding numbers indicated by $w_\delta$ with $\delta=  0, \pm 1, F$, where $w_F$ denotes the winding identified in $F_\perp$. 
Comparison of (e) and (f) shows that our algorithm is able to identify all PCVs and MHVs in the system with a high accuracy.  
A Gaussian blur  filter with 2 grid point width is applied to the data in (a)-(c) and 3 grid point width to the data in (d) to reduce short length scale noise.  }
\end{figure*}

\section{Spin vortices} 
\label{sec:SpinVortices}

In 2D systems vortices often play a dominant role in the phase ordering dynamics.
We  therefore briefly review the structure of single spin vortices in the easy-plane ferromagnetic phase. We write its wave function in polar coordinates with origin taken at  the vortex core.
Sufficiently far from the core the general vortex state vector is of the form
\begin{align}
\bm{\psi}_{V} =  \frac {\sqrt{n_0}} {2} e^{i \sigma_{\phi} \varphi}
\begin{pmatrix}
e^{- i \sigma_{\alpha} \varphi} \sqrt{ 1 - q/q_0}  \\ \sqrt{ 2 (1 + q/q_0)}   \\  e^{i \sigma_{\alpha} \varphi} \sqrt{ 1 - q/q_0} .
\end{pmatrix}.
\end{align}
Here, $\varphi$ is the azimuthal angle, $\sigma_{\phi}$ and $\sigma_{\alpha}$ are integers accounting for the directions of the mass and spin flow around the vortex, respectively.
Different types of spin vortices arise from different combinations of  $\sigma_{\phi}$ and $\sigma_{\alpha}$.
We only discuss elementary vortices given by  $\sigma_{\phi}= 0,\pm 1$ and $\sigma_{\alpha} = \pm 1$ as they will be long-lived configurations in the system.

Polar core vortices (PCVs) exhibit spin circulation ($\sigma_{\alpha} = \pm 1$) but no mass circulation ($\sigma_{\phi}= 0$), and have an unmagnetized (i.e.~polar) core.  
The  two types of PCVs can be distinguished by the phase winding in the order parameter field, i.e.~$F_{\perp} \sim e^{ i \sigma_{\alpha} \varphi}$.
Here we refer to these as positive (p)  $\sigma_{\alpha} = 1$ and negative (n)   $\sigma_{\alpha} = -1$ PCVs, and note that these two types constitute a vortex-antivortex pair and can annihilate.

There are four types of elementary Mermin-Ho vortices (MHVs) which exhibit both mass ($\sigma_{\phi}= \pm 1$) and spin ($\sigma_{\alpha}=\pm1$) circulation, and the vortex core is magnetized. The winding numbers of the $m_{\mathrm{F}}$-th component is given by $w_{m_{\mathrm{F}}}=\sigma_{\phi} - m_{\mathrm{F}}\sigma_{\alpha}$. 
Thus a MHV is characterized by a double winding in either the $m_{\mathrm{F}} = 1$ or $-1$ component, a single winding in $m_{\mathrm{F}} = 0$, and no winding in the remaining component.
We denote the four types of MHVs  as $(\sigma_{\phi} , \sigma_{\alpha}) =$ (p,p), (n,n), (p,n), (n,p) where p $\equiv +1$ and n $\equiv -1$. The spin circulation
$\sigma_{\alpha}$ can be determined from the phase winding of the transverse spin whereas $\sigma_{\phi}$ can directly be inferred from the phase winding of the $m_{\mathrm{F}} = 0$ component. 
This unambiguously characterizes all types of MHVs. 
The two MHVs (p,p) and (n,n) constitute a vortex-antivortex pair, and similarly for  (p,n) and (n,p).

\section{Phase ordering dynamics}
\label{sec:PhaseOrderingDynamics}

\subsection{Numerical methods and parameter quench}
\label{sec:NumericalMethods}

We simulate the phase ordering dynamics starting from the anti-ferromagnetic initial state (\ref{EqAF}). To seed the growth of unstable modes due to the quench, and the subsequent formation of symmetry breaking domains, it is crucial to account for fluctuations beyond mean-field order. 
We do this by adding noise to the initial state according to the truncated Wigner prescription \cite{Blakie2008a, Polkovnikov2010a,Williamson2016a.PhysRevA.94.023608}. 
The time evolution of this initial state is then given by the spin-1 Gross-Pitaevskii equations (GPEs)
\begin{equation} \label{eq:GPE}
i \hbar \partial_t \bm{\psi} =   \left ( - \frac{\hbar^2 \nabla^2}{2M} + q f_z^2 + c_0 n + c_1 \vec{F}\cdot\vec{f} \right) \bm{\psi},
\end{equation}
which are solved by means of a spectral split-step algorithm.
Our quasi-2D simulations are performed  for the case of $c_0 / \lvert c_1 \rvert = 3$, and  $n_0 = 10^4  \, \xi_\mathrm{s}^{-2}$, where $\xi_\mathrm{s} = \hbar/ \sqrt{M q_0}$ is the spin healing length of the system. Each component of the spinor field is represented on a  2D grid of $2048\times2048$ points of spatial extent $l \times l$ $= 1600\,\xi_\mathrm{s} \times  1600\,\xi_\mathrm{s}$, and subject to periodic boundary conditions.

We consider a sudden quench made by setting $q$ to a value in the range $[0,q_0]$  at the start of the simulation.  
In the early time evolution the dynamics is dominated by the growth of unstable modes leading to the formation of transverse magnetization [see, e.g., \cite{Saito2007b.PhysRevA.76.043613,Barnett2011a}]. 
Spin vortices develop between the small magnetized domains that form with a length scale comparable to $\xi_s$. 
Once the (local) ferromagnetic order is established these defects are topologically stabilized and can only decay by mutual annihilation when the appropriate vortex-antivortex pair meets. 
This process is relatively slow compared to the initial growth of local order and dominates the long-time phase ordering dynamics of the system \cite{Bray1994a.AdvPhys.43.357}.
The mean distance between spin vortices is proportional to the average size of the magnetic domains, and is generally taken as the key length scale for the universal phase ordering process.

\subsection{Detection of spin vortices}
\label{sec:DetectionOfSpinVortices}

Compared to the well studied case of quenches from the polar initial state, where only PCVs emerge, our initial state gives rise to a rich ensemble of different vortices. 
An example of the vortex configuration in a subregion of the system is shown in Fig.~\ref{fig:SnapshotSingleRun}.  
This example is taken at a time sufficiently long after the quench such that the average domain size is much larger than the microscopic length scales of the system ($\xi_s$). 

Each spin vortex is located by finding a vortex in the phase field of the transversal spin [see Fig.~\ref{fig:SnapshotSingleRun}(d)].  
The spin vortex type is identified as follows: We count the number and the corresponding winding of vortices occurring in the phase fields of the three  $m_{\mathrm{F}}$ components in a specified detection area around the spin vortex [see Fig.~\ref{fig:SnapshotSingleRun}(a)-(c) and (f)].
The initial detection area is taken to be $3\times3$ grid points.
After extracting all the information from the phase fields we are usually able to unambiguously determine the type of the spin vortex.
We find that some of the spin vortices are stretched (i.e.~the vortices in the different $m_{\mathrm{F}}$ components are spatially separated) due to the heating from the energy released by the quench.
Such vortices cannot be identified if they extend beyond the initial detection area, so we repeat the identification step using a larger detection area in an iterative procedure. 
In each iteration step we increase each side of the detection area by one grid point.
We terminate the algorithm when the spin vortex has been identified or the detection area has grown to include a vortex number exceeding a threshold value.
Note that a small systematic error can arise in our identification analysis when a vortex-antivortex pair is separated by less than  three grid points,  which can for example happen when they are about to annihilate.
The result of the vortex identification analysis is shown in Fig.~\ref{fig:SnapshotSingleRun}(e).
We observe all types of PCVs and MHVs to be present in the system [see \Fig{SnapshotSingleRun}(e) and (f)], and that our algorithm is able to identify them accurately.
In addition to spin vortices, free vortices occur in each of the $m_{\mathrm{F}}$ components and tend to cluster into small groups with the same phase winding.

\begin{figure}
\includegraphics[width = 0.94\columnwidth]{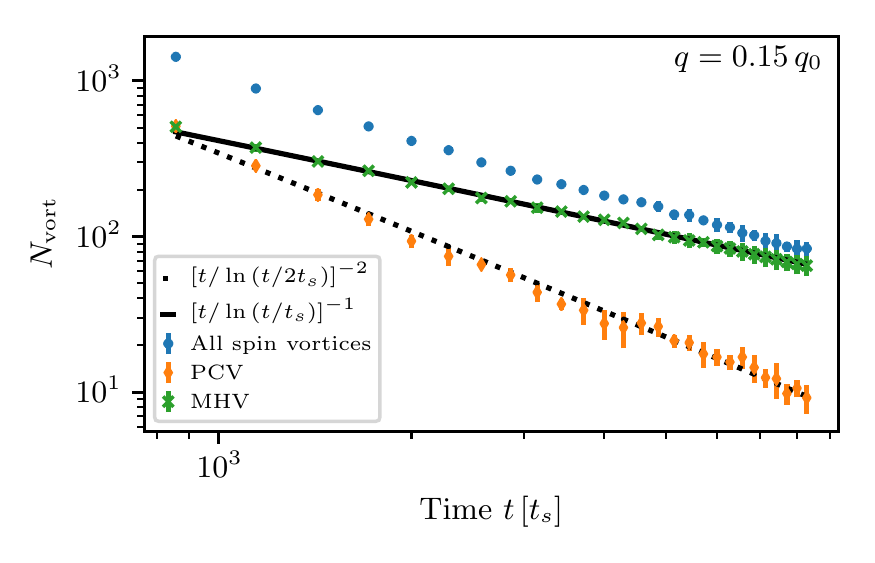}
\caption{\label{fig:VortexCount1} Vortex number $N_{\mathrm{vort}}$ as a function of time within the phase ordering regime for $q=0.15 \, q_0$. The decay of 
PCVs (orange diamonds) and fit Eq.~(\ref{Npcv}) (dotted line). The decay of MHVs (green crosses) and fit Eq.~(\ref{Nmhv}) (solid line). 
Total number of spin vortices (blue dots). 
The data is averaged over 5 trajectories and the error bars correspond to the standard deviation of the results.} 
\end{figure}

\begin{figure}
\includegraphics[width = 0.94\columnwidth]{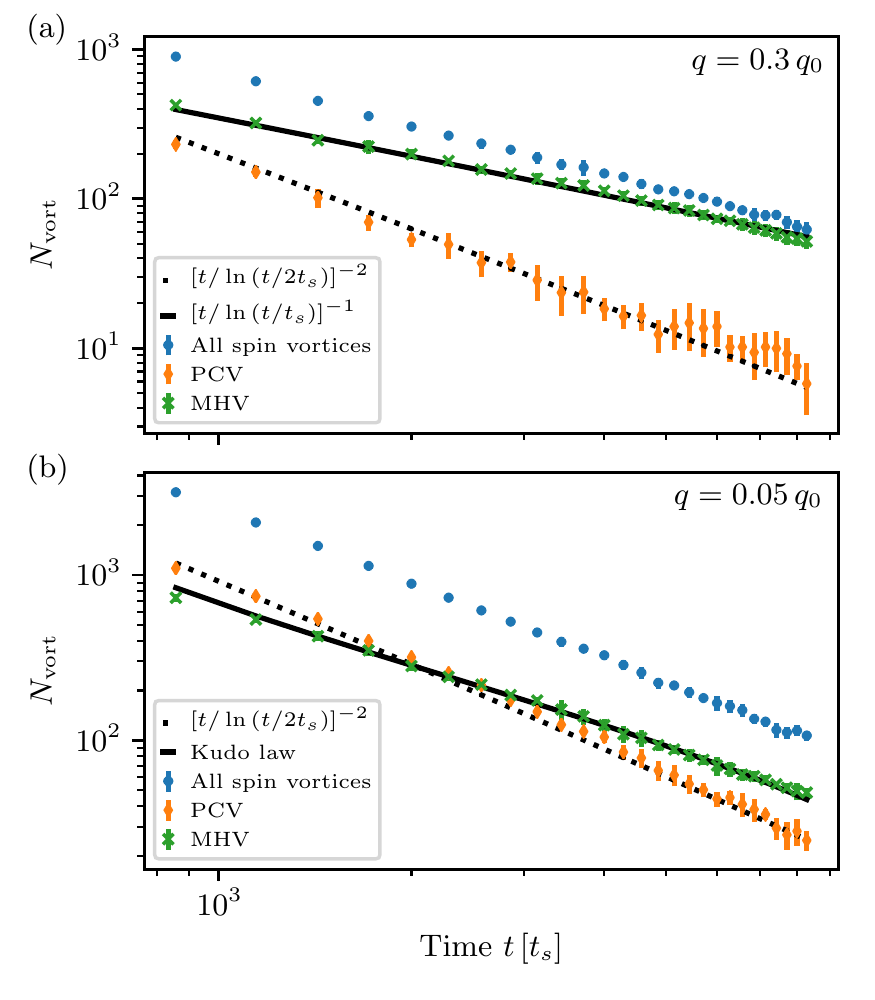}
\caption{\label{fig:VortexCount2} Vortex number $N_{\mathrm{vort}}$ as a function of time within the phase ordering regime for quenches to (a) $q = 0.3 \, q_0$ and (b) $q = 0.05 \, q_0$. The decay of PCVs (orange diamonds) and fit Eq.~(\ref{Npcv}) (dotted lines). The decay of MHVs (green crosses). 
(a) Fit Eq.~(\ref{Nmhv}) (solid line). (b) Fit Kudo decay  law (solid line) according to \cite{Kudo2015a}. 
Total number of spin vortices (blue dots). 
The data is averaged over 5 trajectories and the error bars correspond to the standard deviation of the results. 
} 
\end{figure}

\subsection{Universal decay laws of spin vortices}
\label{sec:UniversalDecayLaws} 
 
To characterize the phase ordering dynamics following the quench we quantify the evolution of the spin vortex number in the system.
Fig.~\ref{fig:VortexCount1} shows the total number of PCVs (orange diamonds) and MHVs (green crosses) at times $850 \, t_s \lesssim  t \lesssim 8300 \, t_s$ for a quench to $q= 0.15 \, q_0$.
There is approximately an equal number of PCVs and MHVs at the earliest time presented however the decay rate of each type of vortex is distinctly different, with the PCVs decaying faster than the MHVs. This leads to
qualitatively different regimes for the phase ordering dynamics: 
As the MHVs become dominant at later times the rate of decay of the total number of spin vortices changes (and hence the magnetic domain growth law)  approaching that of the MHVs.
 
 We quantitatively determine the decay laws for the two types of spin vortices.
The PCV decay is consistent with 
\begin{align}
N_{\mathrm{vort}} \sim \left [t/ \ln \left(t/t_0\right) \right]^{-2},\label{Npcv}
\end{align}
where $t_0$ is a short-time cutoff \cite{Bray2000PhRvL..84.1503B}  [see dotted line in Fig.~\ref{fig:VortexCount1}].
This result agrees with the domain growth law and vortex decay rate found in earlier work on the polar to easy-plane quench where only PCVs emerge \cite{Williamson2016a.PhysRevLett.116.025301}.
The MHVs decay more slowly,
consistent with XY-like scaling \cite{Bray2000PhRvL..84.1503B, Blundell1994, Rutenberg1995b}
\begin{equation}
N_{\mathrm{vort}}\sim \left[t/\ln \left(t/t_0 \right)\right]^{-1},\label{Nmhv}
\end{equation}
[see solid line in Fig.~\ref{fig:VortexCount1}]. 

We have also considered quenches to other values of $q$, and present results for two other cases in  Fig~\ref{fig:VortexCount2}.
These results reveal that the respective decay laws we have identified for the PCVs and MHVs are universal. 
For the quench to $q = 0.05 \, q_0$ we find that we have to modify the XY decay law stated in Eq.~(\ref{Nmhv}) taking into account the possible difference in the number of vortices between the two subclasses of MHVs. 
The decay is well fit \footnote{We fit to Eq.~(39) of \cite{Kudo2015a} with $a= 30957 \pm 1433$, $c= 312\pm 69$ and $d= 17 \pm 2$. We take the parameter $b = l^2 / \xi_s^2 = 2.56 \times 10^6$ as an estimate for the maximal number of vortices. The curve in Fig.~\ref{fig:VortexCount2}(b) is then obtained by numerically inverting  Eq.~(39) for the vortex number.}  by a decay law found  for MHVs in a similar parameter regime ($q/q_0\ll1$, close to the isotropic phase at $q=0$)
by Kudo \textit{et al.}~\cite{Kudo2015a}.  This Kudo decay law is governed by the XY universality class, so this does not indicate a change in the universality of the MHV decay. 

A key feature we observe is that varying the quadratic Zeeman energy we can engineer the proportion of PCVs and MHVs that are present at the start of the phase ordering dynamics. Increasing the quadratic Zeeman energy the ratio of PCVs to MHVs  decreases [see Figs.~\ref{fig:VortexCount1} and \ref{fig:VortexCount2}].

We have also studied quenches for a larger  interaction parameter ratio of $c_0 / \lvert c_1 \rvert = 12$ and find the number and the ratio of spin vortices as well as their decay to be consistent with the results presented above. 

\subsection{Violation of single-length scaling}
\label{sec:ViolationOfSingleLengthScaling}

We find that our system evolution violates  
the dynamic scaling hypothesis, which underlies standard universal phase ordering. The dynamic scaling hypothesis \cite{Bray1994a.AdvPhys.43.357} states that correlation functions of the order parameter collapse (i.e.~become time independent) when spatial coordinates are scaled by the (single) macroscopic length scale $L(t)$. Our system instead has two distinct macroscopic length scales which have different scaling with time: the mean distance between PCVs 
\begin{equation}
\label{eq:LPCV}
L_{\mathrm{PCV}} (t)\sim t/\ln t
\end{equation} 
and the mean distance between MHVs 
\begin{equation}
\label{eq:LMHV}
L_{\mathrm{MHV}} (t)\sim (t/\ln t)^{1/2}.
\end{equation}
Only in the limit of one spin vortex type being much more numerous than the other will pure single length scaling according to the dynamic scaling hypothesis hold.

\begin{figure}
\includegraphics[width = 0.96\columnwidth]{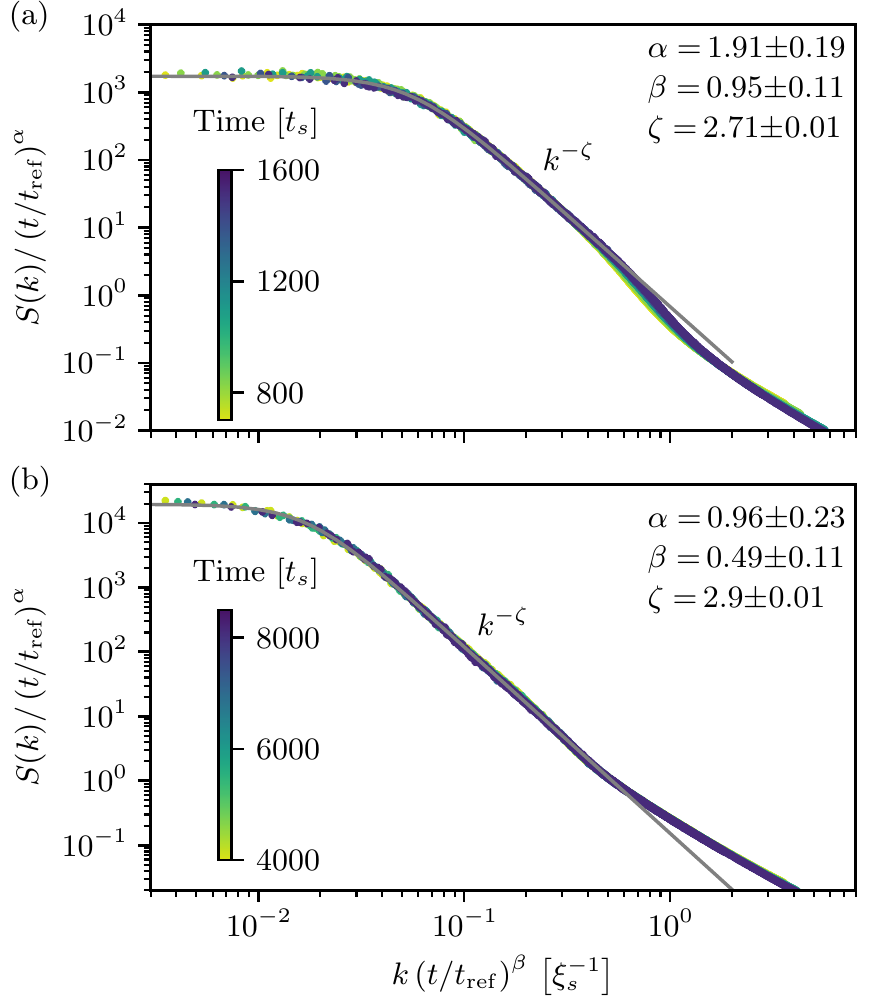}
\caption{\label{fig:StructureFactor}
Universal scaling dynamics of the momentum-space correlation function \eq{StructureFactor} of the transversal spin, $S(k,t)$, according to \Eq{Scaling2} for a quench to $q =0.15 \, q_0$.
(a) Using the scaling exponents $\alpha = 1.91 \pm 0.19$ and $\beta = 0.95 \pm 0.11$ and taking the reference time to be $t_{\mathrm{ref}}= 690 \, t_s$, the data collapses to a universal scaling function for times $650 \, t_s \lesssim t \lesssim 1700 \, t_s$. 
The scaling exponents are consistent with the scaling law obtained for PCVs [see \Eq{LPCV}].
The power-law fall-off of the distribution as $S(k,t) \sim k^{-\zeta}$ is given by $\zeta = 2.71 \pm 0.01$, which differs from the exponent $\zeta = 2.45$ reported for a system containing PCVs only \cite{Williamson2016a.PhysRevLett.116.025301}. 
(b) Using the scaling exponents $\alpha = 0.96 \pm 0.23$ and $\beta = 0.49 \pm 0.11$ and taking the reference time to be $t_{\mathrm{ref}}= 4137 \, t_s$, the data collapses to a universal scaling function  for times  $4100 \, t_s \lesssim t \lesssim 8300 \, t_s$. 
The scaling exponents are consistent with the scaling law obtained for MHVs [see \Eq{LMHV}].
The power-law fall-off of the distribution as $S(k,t) \sim k^{-\zeta}$ is given by $\zeta = 2.90 \pm 0.01$.
All scaling exponents are obtained by means of a least-square fit to the corresponding data
within the infra-red momentum regime below the scale $k_{\mathrm{max}} \xi_s= 0.4$ and  $k_{\mathrm{max}} \xi_s = 0.2$ for the cases (a) and (b) respectively.
The exponent $\zeta$ results from fitting the scaling form $A/[1 +(k/k_L)^{\zeta}]$, with characteristic momentum scale $k_L \sim L(t)^{-1}$, to the above stated infra-red regime of the rescaled data.
The solid grey lines show the best fit of the scaling form.
All data depicted is averaged over 64 trajectories.
} 
\end{figure}

\begin{figure*}
\includegraphics[width = 0.96\textwidth]{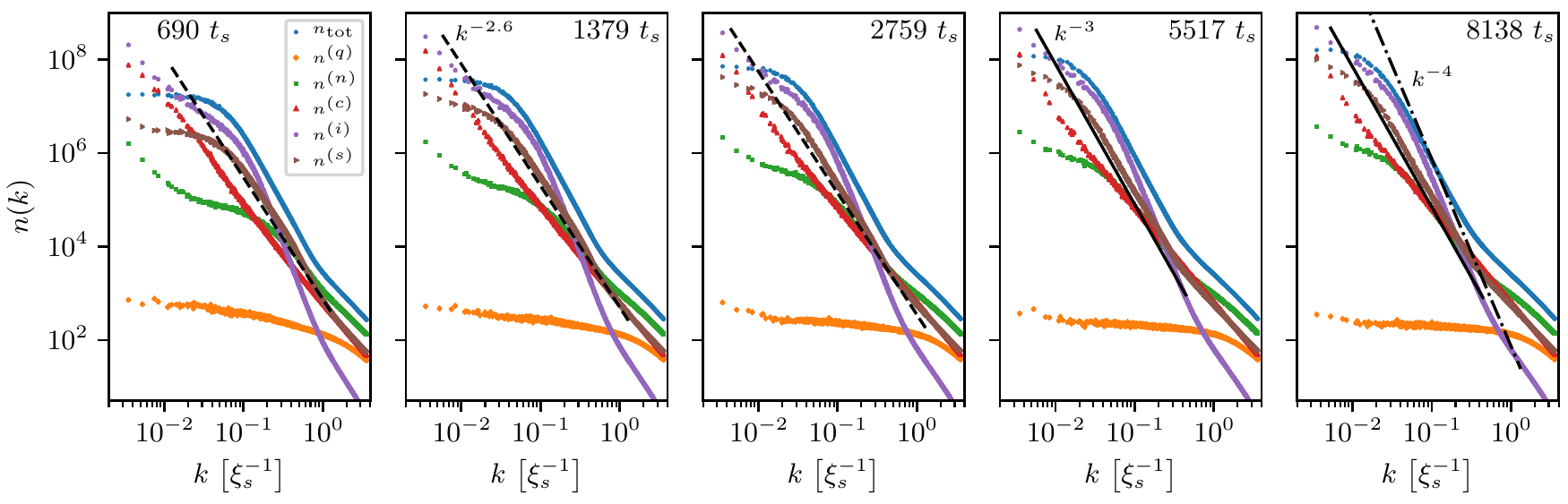}
\caption{\label{fig:HydroDecomposition}
Momentum distributions $n^{(\delta)}(k)$ derived from the hydrodynamic decomposition of the kinetic energy density into $\varepsilon^{(\delta)}(k) = k^{2}n^{(\delta)}(k)/(2M)$ at five different evolution times during the phase ordering process.
The momentum distributions representing the quantum pressure $n^{(q)}$ (orange diamonds), nematic $n^{(n)}$ (green squares), compressible $n^{(c)}$ (red triangles), incompressible $n^{(i)}$ (purple pentagons) and the spin $n^{(s)}$ (brown arrows) parts of the decomposition are compared to the total occupation number $n_{\mathrm{tot}} = \sum_m \lvert \psi_m \rvert^2$ (blue dots).  
See the Appendix for a detailed definition of each of the parts.
The two leftmost panels correspond to the time regime depicted in \Fig{StructureFactor}(a). 
The power-law fall-off of the spin part $n^{(s)}(k) \sim k^{-\zeta}$ with exponent $\zeta \simeq 2.6$ (dashed line) is consistent with the one extracted for the correlation function of the transversal spin.
The two rightmost panels correspond to the time regime depicted in \Fig{StructureFactor}(b). 
Here, the power-law fall-off of the spin part with exponent $\zeta \simeq 3$ (solid line) is also consistent with the one extracted for the correlation function.
This indicates that the spin part plays the dominant role for the shape of the universal scaling function describing the scaling evolution
of the transversal spin.
The incompressible part arising from the vortices in the system shows a power-law behavior with $\zeta \simeq 4$ (dash-dotted line in the rightmost panel) which is consistent with $\zeta = d+2$ predicted for an ensemble of randomly distributed vortices in a $d$-dimensional system \cite{Bray1994a.AdvPhys.43.357}.
} 
\end{figure*}

We verify the above mentioned properties by studying the momentum-space correlation function of the transversal spin
\begin{equation}
\label{eq:StructureFactor}
S(k,t) = \langle \left | f_{\perp} (k,t) \right|^2 \rangle,
\end{equation}
where $f_{\perp} = F_{\perp}/n_0$ and the brackets denote an average over different trajectories of the simulation.
According to the scaling hypothesis, a self-similar evolution of the correlation function, involving a single macroscopic length scale $L(t)$ only, is given by the scaling form
\begin{equation}
\label{eq:Scaling}
S(k,t)  = \left[ L(t)\right]^{\alpha/\beta}f_\mathrm{s}\left( L(t) \, k\right), 
\end{equation}
where $f_s$ is a universal scaling function and $\alpha$, $\beta$ are the corresponding scaling exponents.
As the single length scale $L(t)$ evolves in time according to $L(t) \sim t^{\, \beta}$ we can write \Eq{Scaling} in the more general form
\begin{equation}
\label{eq:Scaling2}
S(k,t) =  \left(t/t_{\mathrm{ref}}\right)^{\alpha}f_\mathrm{s}\left( \left[t/t_{\mathrm{ref}}\right]^{\, \beta} k\right) \,,
\end{equation}
with $t_{\mathrm{ref}}$ being some reference time within the scaling regime. 
If the integral over the correlations is conserved in time within the infra-red momentum regime obeying the scaling evolution, one finds the constraint $\alpha = 2 \beta$ for a two-dimensional system. 

\Fig{StructureFactor} shows the momentum-space correlation function of the transversal spin rescaled according to \Eq{Scaling2} for two different regimes of the time evolution in case of a quench to $q = 0.15 \, q_0$. 
For times $650 \, t_s \lesssim t \lesssim 1700 \, t_s$ the correlation function exhibits approximate scaling with scaling exponents $\alpha = 1.91 \pm 0.19$ and $\beta = 0.95 \pm 0.11$ [see \Fig{StructureFactor}(a)], whereas for times  $4100 \, t_s \lesssim t \lesssim 8300 \, t_s$ we extract scaling exponents $\alpha = 0.96 \pm 0.23$ and $\beta = 0.49 \pm 0.11$ [see \Fig{StructureFactor}(b)].
The exponents result from performing a least-square fit of the data using the reference times $t_{\mathrm{ref}} = 690 \, t_s$ and $t_{\mathrm{ref}} = 4137 \, t_s$, respectively. 
The errors are deduced from the width of the marginal-likelihood functions of the scaling exponents \cite{Orioli:2015dxa}.
The extracted scaling exponents are consistent with the integral of the correlation function being conserved in time within the infra-red scaling regime as we find $\alpha \approx 2 \beta$.
Note that we do not take into account a possible logarithmic correction entering the scaling forms \eq{Scaling} and \eq{Scaling2} in our analysis. 
As the time window considered for the scaling analysis of the correlation function is comparatively small we expect the effects of logarithmic 
corrections to not be detectable within the error of the extraction method. 

We clearly observe two distinct scaling regimes for the time evolution of our spin-1 system. 
The scaling exponents for the early stage of the phase ordering are consistent with the scaling law obtained for PCVs [see \Eq{LPCV}], whereas we find good agreement
with the scaling of MHVs [see \Eq{LMHV}] within the late-time regime. 
For times $1700 \, t_s \lesssim t \lesssim 4100 \, t_s$ we are not able to collapse the data with a single set of exponents $\alpha$, $\beta$. 
This indicates the violation of single-length scaling in the system.
While the decay of each of the underlying spin vortices obeys a universal scaling law during the whole process of phase ordering,
the correlation function measuring the evolution of the order parameter does not.
Nonetheless, in the late-time regime where MHVs are much more numerous than PCVs [see \Fig{VortexCount1}] we find the phase ordering process to be well described by a single length scale only corresponding to the decay of MHVs [see \Fig{StructureFactor}(b)].
Due to the fast decay of PCVs as compared to MHVs, the phase ordering is dominated by the PCV scaling law at early stages [see \Fig{StructureFactor}(a)],
although there is an approximate equal number of PCVs and MHVs in the system [see \Fig{VortexCount1}]. 

However, the phase ordering for evolution times $t \lesssim 1700 \, t_s$ is not purely characterized by the dynamics and the properties of PCVs. 
This becomes visible when investigating the scaling function associated with the scaling evolution in \Fig{StructureFactor}(a). 
The momentum-space correlation function of the transversal spin shows a plateau below the characteristic momentum scale $k_L \sim L(t)^{-1}$, followed by a power-law fall-off $S(k,t) \sim k^{-\zeta}$.
Using all rescaled data we determine the exponent $\zeta$ by means of fitting the scaling form $A/[1 +(k/k_L)^{\zeta}]$ to the infra-red momentum regime.
We extract an exponent $\zeta = 2.71 \pm 0.01$ from the fit, which is considerably larger than the exponent $\zeta = 2.45$, which was found for a system containing PCVs only \cite{Williamson2016a.PhysRevLett.116.025301}. 
We expect the deviation to arise from the approximately equal number of MHVs being present in the system.
In contrast, in the late-time regime, we extract an exponent of $\zeta = 2.90 \pm 0.01$. This indicates that not only the scaling exponents but also the shape of the scaling function differs for the two types of spin vortices causing the associated universal dynamics to belong to clearly 
distinct universality classes.

We emphasize that the power-law fall-off of the correlation function of the transversal spin does not characterize the flow fields induced by the vortices detected in the system.
For an ensemble of randomly distributed vortices one generally expects a steeper power-law fall-off of the order parameter correlation function
with exponent $\zeta = d +2$ \cite{Bray1994a.AdvPhys.43.357, Nowak:2011sk, Nowak:2013juc}.

\Fig{HydroDecomposition} shows the momentum distributions $n^{(\delta)}(k)$ derived from the hydrodynamic decomposition of the kinetic energy density as defined in the Appendix. 
Note in particular the (purple pentagons) curve $n^{(i)}(k)$ which depicts the contribution of the incompressible, i.e., the divergence-free part of the 
velocity field to the kinetic energy spectrum.
This part arises from both, spin and free vortices [c.f.~\Fig{SnapshotSingleRun}(f)] and falls off as $n^{(i)}(k)  \sim k^{-\zeta}$ with $\zeta \simeq 4$ during the late-time regime of the phase ordering.
We remark that for the early stage we observe a slightly steeper power-law consistent with $\zeta \simeq 4.3.$

As a result, while the total kinetic energy spectrum is dominated by the contribution from the incompressible part within the infra-red momentum region below $k \xi_s \lesssim  0.2 $, the power-law fall-off of the spin correlator $S(k,t)$ is closer to that of the spin part of the decomposition, for which we find an exponent consistent with $\zeta \simeq 2.6$ for times  $650 \, t_s \lesssim t \lesssim 1700 \, t_s$, and an exponent $\zeta \simeq 3$ in the late-time regime $4100 \, t_s \lesssim t \lesssim 8300 \, t_s$ [see dashed and solid lines respectively in \Fig{HydroDecomposition}]. 
Both exponents can be related to the surface structure of the transversal spin domains present in the system.
For domains in a $d$-dimensional system one generally expects a power law behavior of the associated momentum-space correlator with exponent $\zeta = -2d+d_{\mathrm{s}}$, where $d_{\mathrm{s}}$ denotes the surface fractal dimension \cite{Oh1999}. 
For smooth surfaces the fractal dimension is $d_{\mathrm{s}}= d-1$ \cite{Oh1999}, which results in an exponent $\zeta = 3$. 
Hence, the late-time scaling regime dominated by the annihilation of MHVs can be interpreted in terms of the spin domains having a
rather smooth surface structure.
However, within the early stage of the phase ordering the scaling is consistent with the spin domains having a fractal surface structure with fractal dimension $d_{\mathrm{s}}\approx 1.4$. 
Note that a fractal dimension of this size has been found for the phase ordering process involving PCVs only \cite{Williamson2016a.PhysRevLett.116.025301}.
Our results thus seem to indicate that each type of spin vortex is accompanied by a specific surface structure of the attached spin domain boundaries.

\section{Conclusion and Outlook} 
\label{sec:Conclusion}
In this work we studied the phase ordering dynamics of a system quenched into the easy-plane ferromagnetic phase.
Our choice of novel initial condition allows both PCVs and MHVs to form during the quench and subsequently we find that both types of spin vortices play a crucial role in the phase ordering. 
Because the two types of vortices have different decay laws, the standard (i.e.~single macroscopic length scale) dynamic scaling hypothesis cannot hold for this system. 
We find that the ratio of PCVs and MHVs produced can be varied by quenching to different $q$. 
The subsequent decay of each type of spin vortex appears universal. 
We believe this presents an extension of the dynamic scaling hypothesis to systems supporting multiple defects relevant to the order parameter.

A feature of our system is that it can be realized in experiments with spinor Bose-Einstein condensates. The initial state production and tuning of $q$ are common place experimental manipulations. The main challenges lie in producing a large quasi-2D system, ideally in a flat bottomed trap, and subsequently monitoring the evolution for long time scales.

The initial condition considered here is a $\pi/2$-spin rotation of that studied in earlier work, yet the equilibration dynamics proceeds in a different manner involving a new class of topological defects. 
An interesting future direction is to vary the spin rotation angle continuously to produce a family of initial states to explore the crossover between the two different transient ordering processes.

\begin{acknowledgments}
The authors thank L.~M.~Symes, L.~A.~Williamson and the SynQS-Team in Heidelberg for discussions and collaboration on the topics described here. 
This work was supported by the Horizon-2020 programme of the EU (ERC Adv.~Grant EntangleGen, Project-ID 694561),  by DFG (SFB 1225 ISOQUANT), by DAAD (No.~57381316), and by Heidelberg University (CQD).
C.-M.S. thanks the Dodd-Walls Centre, University of Otago, New Zealand, for hospitality and support. 
This work was performed on the bwForCluster MLS\&WISO supported by the state of Baden-W\"urttemberg through bwHPC.
\end{acknowledgments}

\begin{appendix}
\begin{center}
\textbf{APPENDIX}
\end{center}
\setcounter{equation}{0}
\setcounter{table}{0}
\makeatletter

\begin{figure*}
\includegraphics[width = 0.96\textwidth]{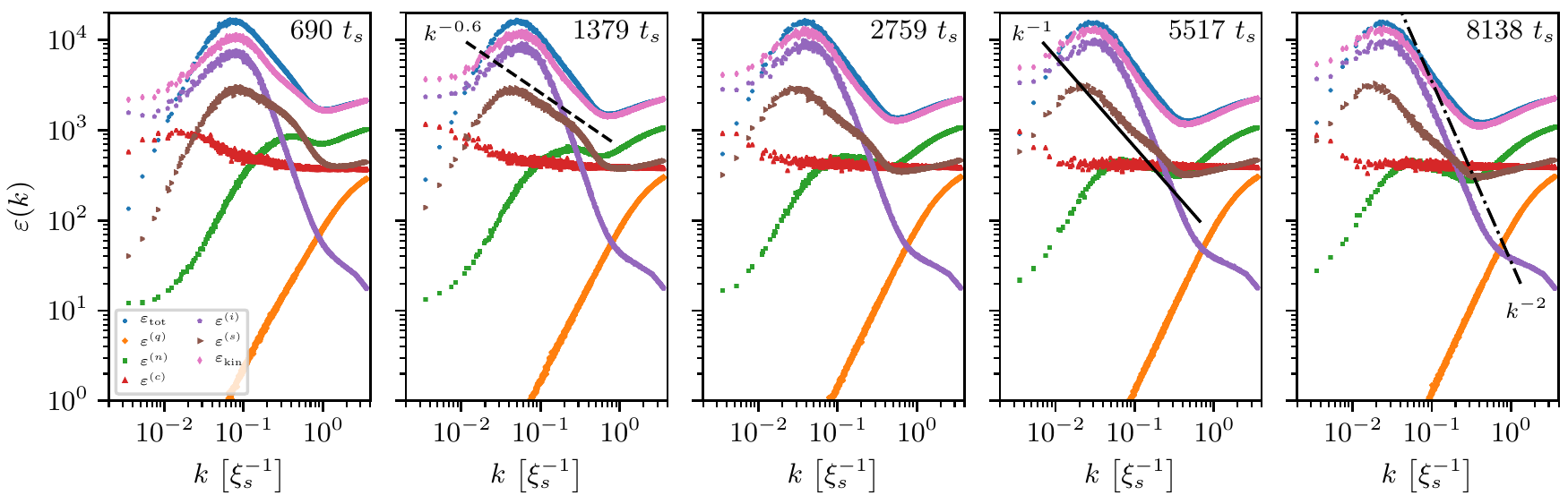}
\caption{\label{fig:HydroDecompositionEnergy}
Hydrodynamic decomposition of the kinetic energy density into $\varepsilon^{(\delta)}(k)$ for the same evolution times as in \Fig{HydroDecomposition}.
The contributions representing the quantum pressure $\varepsilon^{(q)}$ (orange diamonds), nematic $\varepsilon^{(n)}$ (green squares), compressible $\varepsilon^{(c)}$ (red triangles), incompressible $\varepsilon^{(i)}$ (purple pentagons) and the spin $\varepsilon^{(s)}$ (brown arrows) parts of the decomposition are compared to $\varepsilon_{\mathrm{tot}} (k) = k^2 n_{\mathrm{tot}}(k)/(2M)$ (blue dots).  
The sum of all parts of the decomposition $\varepsilon_{\mathrm{kin}} = \sum_\delta \varepsilon^{(\delta)}$ (pink thin diamonds) reveals expected deviations between the hydrodynamic decomposition and  $\varepsilon_{\mathrm{tot}}$ for infra-red momenta \cite{Nowak:2011sk}.
While the spin part is characterized by a single power-law according to $\varepsilon^{(s)} (k) \sim k^{-\zeta + 2}$ with $\zeta \simeq 2.6$ (dashed line)
in the early stage of the phase ordering, a bimodal distribution with a steeper power-law fall-off consistent with $\zeta \simeq 3$ (solid line) within the low momentum region arises in the late-time regime.
The incompressible part shows a power-law behavior consistent with $\zeta \simeq 4$ (dash-dotted line).
We remark that the power-law fall-off of $\varepsilon_{\mathrm{tot}}$ arises from contributions of both, the incompressible and the spin part.
Hence it neither gives direct access to the vortices characterizing the scaling evolution nor to the surface structure of the spin domains determining the shape of the universal scaling function of the transversal spin correlator.
} 
\end{figure*}

\section{Hydrodynamic decomposition}
\label{app:HydroDecomposition}

In this Appendix we provide a brief definition of the hydrodynamic decomposition of the spin-1 Bose gas. 
For details see also \cite{Yukawa2012a.PhysRevA.86.063614, Mikheev:2018adp}.
For simplicity of the expressions we use units of $\hbar =1$.
In a hydrodynamic formulation \cite{Yukawa2012a.PhysRevA.86.063614}, the spin-1 system is described by the total density $\rho$, the spin vector $f_{\mu}$ and the nematic tensor $n_{\mu\nu}$, 
\begin{align}
  \rho 
  &=  \sum_{m} \psi_m^{\dagger}  \psi_m\,
  \\
  f_{\mu} 
  &= \frac{1}{\rho} \sum_{m, m^\prime} \psi_m^{\dagger} \left( \mathrm{f}_{\mu} \right)_{m m^\prime} \psi_{m^\prime} \,,    
  \label{eq:spinop}
  \\
  n_{\mu \nu} 
  &= \frac{1}{\rho} \sum_{m, m^\prime} \psi_m^{\dagger} \left( \mathrm{n}_{\mu \nu} \right)_{m m^\prime} \psi_{m^\prime}\,,
  \label{eq:nemop}
\end{align}
$\mu=x,y,z$, with $\mathrm{f}_{\mu}$ being the spin-1 matrices in the fundamental representation, and the nematic or quadrupole tensor representation $\mathrm{n}_{\mu \nu} =  ( \mathrm{f}_{\mu} \mathrm{f}_{\nu} + \mathrm{f}_{\nu} \mathrm{f}_{\mu} )/2$.
The superfluid velocity field $\mathbf{v}$ is then given by
\begin{equation}
  \mathbf{v}
  = \frac{-i}{2 M \rho} \sum_m\left[ \psi_m^{\dagger}  \left( \nabla \psi_m \right) 
                                                        - \left(\nabla \psi_m^{\dagger} \right) \psi_m \right]\,.
 \end{equation}
For expressing the hydrodynamic energy it is useful to define the generalized velocities, corresponding to the quantum-pressure (q), the spin (s), the nematic (n), the incompressible (i) and compressible (c) parts,
\begin{align}
\vec{w}^{(q)} &= M^{-1}\nabla \sqrt{\rho}\,,
&\vec{w}^{(i,c)} =&\ \sqrt{\rho}\, \vec{v}^{(i,c)} \,.
\nonumber\\
\vec{w}_{\mu}^{(s)} &= (2M)^{-1}{\sqrt{\rho}}\,\nabla f_{\mu}\,,
&\vec{w}_{\mu \nu}^{(n)} =&\ (2M)^{-1}{\sqrt{2\rho}}\,\nabla n_{\mu \nu}\,,
\end{align}
Here, $\vec{v}^{(i,c)}$ are obtained by a Helmholtz decomposition of $\vec{v} = \vec{v}^{(i)} + \vec{v}^{(c)}$, with the incompressible part having a vanishing divergence,  $\nabla \cdot\vec{v}^{(i)} = 0$,  and the compressible part a vanishing curl, $\nabla \times\vec{v}^{(c)} = 0$.

Using the hydrodynamic variables we can express the energy as 
\begin{equation}
E = E_\mathrm{kin}
+  \int d^2 x \,  \left [ \frac{c_0}{2} \rho^2 + \frac{c_1}{2} \rho^2 f_{\mu}^2 + q \rho n_{zz} \right],
\end{equation}
where the kinetic part  reads
\begin{align}
E_\mathrm{kin} 
=&  \frac{1}{2 M}  \int d^2 x\left[ 
\left( \nabla \sqrt{\rho} \right)^2
+ \frac{\rho}{4} \left( \nabla f_{\mu} \right)^2 
+ \frac{\rho}{2} \left( \nabla n_{\mu \nu} \right)^2 
\right] \nonumber \\
&+\, \frac{M}{2} \int d^2 x \, \rho \vec{v}^2 \,.
\end{align}
Hence, in Fourier space, the kinetic-energy spectrum is given by the correlation functions of the generalized velocities
\begin{equation}
\varepsilon_\mathrm{kin}(k) = \varepsilon^{(q)} (k) + \varepsilon^{(c)} (k) + \varepsilon^{(i)} (k)+ \varepsilon^{(s)} (k) + \varepsilon^{(n)} (k)\,,
\end{equation}
averaged over the orientation of the momentum vector, 
\begin{align}
\varepsilon^{(\delta)} (k) & = \frac{M}{2} \int d \Omega_{\mathbf{k}} \langle \lvert  \vec{w}^{(\delta)} (\mathbf{k}) \rvert^2 \rangle\,,\quad (\delta=q,i,c)
\label{eq:epsq}
\\
\varepsilon^{(s)} (k)  &= \frac{M}{2} \int d \Omega_{\mathbf{k}} \langle   \vec{w}_{\mu} ^{(s)} (\mathbf{k}) \cdot \vec{w}_{\mu} ^{(s)} (\mathbf{k})  \rangle\,,
\label{eq:epss}
\\
\varepsilon^{(n)} (k)  &= \frac{M}{2} \int d \Omega_{\mathbf{k}} \langle   \vec{w}_{\mu \nu} ^{(n)} (\mathbf{k}) \cdot \vec{w}_{\mu \nu} ^{(n)} (\mathbf{k})  \rangle\,,
\label{eq:epsn}
\end{align}
where Einstein's summation convention is implied.
The respective total energies are obtained as $E^{(\delta)}=\int d^2 x \int dk\, k \,  \varepsilon^{(\delta)}(k)$.
The spectrum of the kinetic energy can then be used to calculate corresponding occupation numbers using the relation
\begin{equation}
n^{(\delta)} (k)  = 2M\,k^{-2} \varepsilon^ {(\delta)} (k)\,,
\end{equation}
where $\delta = q, i, c, s, n$.
The total occupation number is then approximately given by 
\begin{equation}
n_\mathrm{tot} (k)  \approx \sum_\delta n^{(\delta)}(k)=  2M\,k^{-2} \varepsilon_\mathrm{kin}  (k)\,.
\end{equation}
We remark that the total occupation number $n_{\mathrm{tot}}$ and $2M\,k^{-2} \varepsilon_\mathrm{kin} (k)$ deviate from each other 
in the regime of infra-red momenta $k \xi_s \lesssim 1$ due to additional contributions from four-point correlations of the fundamental fields contributing to the kinetic energy density, see, e.g., Ref. \cite{Nowak:2011sk}.

\vspace{0.2cm}

\section{Spectrum of the kinetic energy}

In this Appendix we briefly  discuss the spectrum of the kinetic energy of our spin-1 system to give additional insights to the results presented in \Fig{HydroDecomposition} in the main text. 

\Fig{HydroDecompositionEnergy} shows the hydrodynamic decomposition of the kinetic energy density as defined in \App{HydroDecomposition} during the phase ordering process described in the main text.
For the early stage of the phase ordering the spin part of the decomposition exhibits a power-law behavior according to $\varepsilon^{(s)} (k)\sim k^{-\zeta +2 }$ with $\zeta \simeq 2.6$.
In the late-time regime, the single power-law transitions into a bimodal distribution characterized  by 
a steeper power-law fall-off consistent with $\zeta \simeq 3$ in the low momentum region while the high momentum region is still characterized by an exponent $\zeta \simeq 2.6$.
In contrast, for the incompressible part of the kinetic energy, $\varepsilon^{(i)} (k)$,  a single power-law consistent with $\zeta \simeq 4$ prevails throughout the whole evolution. 
Note that the power-law fall-off of the total kinetic energy $\varepsilon_{\mathrm{tot}}(k) = k^2 n_{\mathrm{tot}}(k)/(2M)$ results from both,
the incompressible and the spin part such that it neither provides direct access to the vortices characterizing the scaling evolution nor to the surface structure of the spin domains determining the shape of the universal scaling function of the order parameter correlator.

\end{appendix}


%


\end{document}